\documentclass{elsart}
\usepackage{amsmath,amsfonts,amssymb,epsfig,color}

\newcommand{\spq}[1]{\ensuremath{\mathfrak{sp}_q( #1 )}}

\newcommand{\suq}[2]{\ensuremath{\mathfrak{su}_q^{ #1 }( #2 )}}

\begin{document}

\begin{frontmatter}

\title{On the physical significance of $q$-deformation to isovector
pairing interactions in nuclei}
\author{K.D.~Sviratcheva$^1$, C.~Bahri$^1$, A.I.~Georgieva$^{1,2}$,
J.P.~Draayer$^1$}
\address{$^1$Department of Physics and Astronomy, Louisiana State
University, Baton Rouge, LA 70803, USA \\
$^2$Institute of Nuclear Research and Nuclear Energy,
Bulgarian Academy of Sciences, Sofia 1784, Bulgaria}

\date{\today}

\begin{abstract}
The quantum deformation concept is applied to a study of isovector pairing
correlations in nuclei of the mass $40 \leq A \leq 100$ region.  While the
non-deformed ($q \rightarrow 1$) limit of the theory provides a reasonable
global estimate for strength parameters of the pairing interaction, the results
show that the $q$-deformation plays a significant role in
understanding higher-order effects in the interaction.
\end{abstract}


\end{frontmatter}

\noindent
\textit{1.~Introduction} ---
In addition to purely mathematical examinations of quantum algebraic concepts
(see e.g. \cite{Feng01}), studies focused
on quantum (or $q$-) deformation \cite{Jimbo,Drinfeld,Fad88} 
in various fields of physics has been the focus of considerable attention in
recent years. Studies of interest include applications in string/brane theory,
conformal field theory, statistical mechanics, and metal clusters
\cite{Paw01,Sli01,Alg02,Bon00}. A feature of any quantum theory is 
that in the $q
\rightarrow 1$ limit one recovers the ``classical'' (non-deformed) results.
\newline 
The earliest applications of $q$-deformation in nuclear physics were related to
a description of rotational bands in axially deformed nuclei
\cite{RRS90}
using the second order Casimir operator of $SU_{q}(2)$. Although optimum values
for the deformation parameter did not differ much from the ``classical'' $q
\rightarrow 1$ limit, an overall improved fit to the experimental
excitation energies was achieved with $q \ne 1$.
Nevertheless, such applications can contradict the physical 
interpretation of the
generators of the algebra when the generators are associated with
\textit{fundamental} symmetries of the system \cite{BNW93}.  In 
particular, when
one $q$-deforms $SO(3) \sim SU(2)$, rotational invariant of the system is
compromised.
But the situation can be very different if one considers the algebra
of a many-body interaction. In such scenarios, the $q$-deformation accounts for
non-linear contributions of higher-order interactions
\cite{Sharma,SharmaSh} without  affecting physical
observables associated with a subset of the generators of the algebra that
remain unchanged by the deformation. This approach also
allows one to construct $q$-deformed nuclear  Hamiltonians with
exact solutions. By considering a $q$-deformed generalization of some nuclear
structure models, the role of the deformation can be explored by comparing the
``classical'' and $q$-deformed results with the experimental data.
\newline 
It is well-known that effective interactions in nuclei are dominated by pairing
and quadrupole terms.  The former accounts for the formation of fermion pairs
that give rise to a superconducting/pairing gap in nuclear spectra, and the
latter is responsible for the strong enhanced electric quadrupole 
transitions in
collective rotational bands.
Indeed, within the framework of the harmonic oscillator shell-model with
degenerate single-particle energies, these two limiting cases have a very clear
algebraic structure in the sense that the spectra exhibit a dynamical symmetry.
In the pairing limit (appropriate for near closed-shell nuclei), the 
Kerman-Klein
quasi-spin $SU(2)$ group \cite{Ker} together with its dual, the unitary
symplectic group $Sp(2\Omega)$ \cite{Rac,Flowers}, where $2\Omega$ is equal to
the degeneracy of the shell, allows one to introduce the seniority quantum number
that can be used to classify the spectra. On the other hand, in the quadrupole
limit the symplectic group $Sp(6,\mathbb{R})$ \cite{Ros} governs a
shape-determined dynamics.
\newline 
\indent The inclusion of a proton-neutron isovector (isospin $T=1$) pairing
interaction enlarges the group structure from the Kerman-Klein $SU(2) \sim
Sp(2)$ to $SO(5)
\sim Sp(4)$ \cite{Helmers,Szpikowski,Hecht,EngelLV96}. And a further
extension to the $SO(8)$ group \cite{Pang69} allows one to also take the
competing isoscalar ($T=0$) proton-neutron pairing mode into account
\cite{EngelPSVD97,PalDJ01}. In this letter we focus on isovector pairing
correlations in nuclei and, in addition, include a proton-neutron
isoscalar term in the interaction that is diagonal in an isospin basis. The
simple two-body
$Sp(4)$ pairing model gives a very reasonable estimate for the nuclear
interaction strength, which we assume to be constant for all nuclei within a
major shell. 
The new feature reported on in this
article is an extension of the theory to include non-linear deviations from the
pairing solution as realized through a $q$-deformation of the underlying $Sp(4)$
symmetry group.
\newline 
Our results show that the $q$-deformation parameter
is decoupled from the interaction strength.  In addition, since the $q \ne
1$ results are uniformly superior to those of the non-deformed limit, 
the results
also suggest that the deformation has physical significance over-and-above the
simple pairing gap concept, extending to the very nature of the nuclear
interaction itself. In short, our results suggest that the $q$-deformation has
physical significance beyond what can be achieved by simply tweaking the 
parameters of a
two-body interaction. The results also underscore the need for additional
studies to achieve a more comprehensive understanding of $q$-deformation in 
nuclear physics.

\noindent\textit{2.~The $q$-deformed pairing model} ---
The $\spq{4}$ deformed algebra \cite{Hayashi90,Sel95,fermRealSp4} is
constructed
in terms of $q$-deformed creation and annihilation operators,
$\alpha^\dagger_{jm\sigma}$ and $\alpha_{jm\sigma}$, each of which creates and
annihilates a nucleon with isospin $\sigma$ ($+\half$ for proton, $-\half$ for
neutron) in a single-particle state of total angular momentum $j$
(half-integer)
with projection $m$ along the $z$-axis.  The $q$-deformed fermion operators are
defined through their anticommutation relations, 
$\{ \alpha_{jm\sigma}, \alpha^\dagger_{jn\sigma}
				\}_{\pm 1} =
    q^{\pm \frac{N_{2\sigma}}{2\Omega}}\delta _{mn}, \
    \{ \alpha_{jm\sigma}, \alpha^\dagger_{kn\tau} \} \textstyle{=}
    0\ (j\ne k,\ \sigma \ne \tau )$, and $
    \{ \alpha^{\dagger}_{jm
    \sigma}, \alpha^{\dagger}_{k n\tau} \} \textstyle{=}\{ \alpha_{jm
    \sigma}, \alpha_{k n\tau} \} \textstyle{=} 0$ \cite{fermRealSp4},
where the $q$-anticommutator is $\{A,B\}_\kappa = AB + q^\kappa BA$ and
$N_{2\sigma }\textstyle{=}\left( \sum_{jm} \alpha^\dagger_{jm\sigma}
    \alpha_{jm\sigma}\right)_{q=1}$. In a model with degenerate single-particle
levels, the dimension of the space for given $\sigma$ is $2\Omega = \sum_j
(2j+1)$. The non-deformed (``classical'') version of the theory is obtained in
the limit $q\rightarrow 1$. The generators of $Sp_q(4)$ are constructed as a
bilinear product of fermion operators coupled to total angular momentum and
parity $J^\pi=0^+$, 
\begin{align}
    T_\pm &= \frac{1}{\sqrt{2\Omega}} \sum_{jm} \alpha^\dagger_{jm,\pm \half}
    \alpha_{jm,\mp \half}, \nonumber \\
    A^\dagger_\mu &= \frac{1}{\sqrt{2\Omega (1+\delta_{\sigma\tau})}}
    \sum_{jm} (-1)^{j-m} 
\alpha_{jm,\sigma}^\dagger\alpha_{j,-m,\tau}^\dagger, \label{genSpq} \\
    A_\mu &= (A^\dagger_\mu)^\dagger,\ \mu=\sigma+\tau \nonumber
\end{align}
in addition to the proton (neutron) number operators $N_{\pm 1}$, which remain
undeformed. These
generators of the Cartan subalgebra can also be realized in terms of the isospin
projection operator
$T_0=\half(N_{+1} - N_{-1})$ and the total nucleon number operator
$N=N_{+1} + N_{-1}$. The $\spq{4}$ algebra contains four
distinct
$\suq{}{2}$ $q$-deformed subalgebras (see table~\ref{tab:su2}).
The commutation relations between the generators 
are symmetric with respect to the exchange 
$q\leftrightarrow q^{-1}$ \cite{fermRealSp4}.
\begin{table}[e]
\caption{Realizations of various $\suq{ }{2} \subset \spq{4}$.}
\begin{tabular}{c|cc}
\hline \hline
symmetry ($\mu $) &$\suq{\mu}{2}$ &
\hspace{0.1in}$\mathfrak{u}^\mu(1)$\hspace{0.1in}\\
\hline
$pp$ pairs ($+$) & $A^\dagger_{+1}, \half (N_{+1}-\Omega),A_{+1}$
& $N_{-1}$\\
$nn$ pairs ($-$) & $A^\dagger_{-1}, \half (N_{-1}-\Omega),A_{-1}$
& $N_{+1}$\\
$pn$ pairs (0)   & $A^\dagger _0,\half N - \Omega, A_0$
& $T_0$ \\
isospin ($T$)    & $T_+, T_0, T_-$
& $N$ \\
\hline
\end{tabular}
\label{tab:su2}
\end{table}
\newline 
As for the microscopic ``classical'' approach \cite{KleinMarshalek}, the
most general Hamiltonian of a system with $Sp_q(4)$ dynamical symmetry and
conserved 
proton and neutron particle numbers can be expressed in
terms of the generators (\ref{genSpq}) \cite{Svi02}
\begin{align}
H_q &=-\varepsilon _q N - F_q (A^\dagger _{+1} A_{+1} + A^\dagger _{-1} A_{-1})
- G_q A^\dagger_0 A_0 \nonumber \\
&- \textstyle{2C_q {([\frac{N}{2}-\Omega
]^{*2}-[\Omega ]^{ *2})}
- D_q {[T_0]^{*2}} -\frac{E_q}{2} \left(\{T_+,T_-\}-\left[
\frac{N}{2\Omega}\right]
\right)},
\label{qH}
\end{align}
where $[X]^{*2}=\Omega \left[\frac{X}{2\Omega }\right] \left(\left[
X+1\right] _{\frac{1}{2\Omega }}+\left[ X-1\right] _{\frac{1}{2\Omega
}}\right)
\stackrel{q\rightarrow 1}{\rightarrow }X^2$ and by
definition $[X]_{\kappa}=\frac{q^{\kappa X}-q^{-\kappa X}}{
q^{\kappa}-q^{-\kappa}} \stackrel{q\rightarrow 1}{\rightarrow }X$.
In principle, the 
parameters $\gamma _q=\{
\varepsilon _q, F_q, G_q,C_q, D_q, E_q
\}$ can be different from their non-deformed counterparts $\gamma
=\{\varepsilon , F, G, C, D, E \}$.
The model describes the motion of $N_+$ valence protons and $N_-$ valence
neutrons in the mean-field of a doubly-magic nuclear core. The
basis states are constructed by the action of the pair-creation operators
$A^\dagger _{0,\pm 1}$, on the vacuum. The quantum numbers used to
specify the basis set $|n_1,n_0,n_{-1})$ count the number of
$pp$, $pn$ and
$nn$ pairs, respectively, and are associated with the corresponding limits of
$\spq{4}$ (see table \ref{tab:su2}). The basis vectors model $0^+$
states with dominant isovector pair correlations. In the mass $40 \le A
\le 100$ region they can be used to
describe $0^+$ ground states of even-$A$ nuclei and
the lowest isobaric analog
$0^+$ state in  the odd-odd nuclei with a $J \ne 0$ ground state. 
\newline
From a ``classical'' perspective, the deformation introduces higher-order,
many-body terms into a theory that starts with only one-body and two-body
interactions, the latter including an isovector  pairing interaction
(parameters $F,~G$) together with a proton-neutron isoscalar force 
diagonal in an isospin basis. 
The way in which the higher-order effects enter into the theory is
governed by the
$[X]$ form. Since $\varkappa$ and $q$ are related to one another,
$q\textstyle{=}e^{\varkappa }$, everything is tied to the deformation with $[X]
\textstyle{=}\frac{\sinh{(\varkappa  X )}}{\sinh{(\varkappa )}}\textstyle{=}
X(1\textstyle{+} \varkappa ^{2} \frac{X^{2}-1}{6} \textstyle{+}
\varkappa ^{4} \frac{3X^{4}-10X^{2}+7}{360}
\textstyle{+} ...) \textstyle{\rightarrow} \allowbreak X $ 
in the $\varkappa \textstyle{\rightarrow} 0$ 
($q\textstyle{\rightarrow} 1$) limit.
The
deformation is applied to the same region of nuclei where the ``classical"
model has already proven to provide for a reasonable description of the $0^+$
states under consideration. Thus the $q$-deformation does not remedy the
non-deformed model but complementarily can improve it.

\begin{figure}[e]
\centerline{\hbox{\epsfig{figure=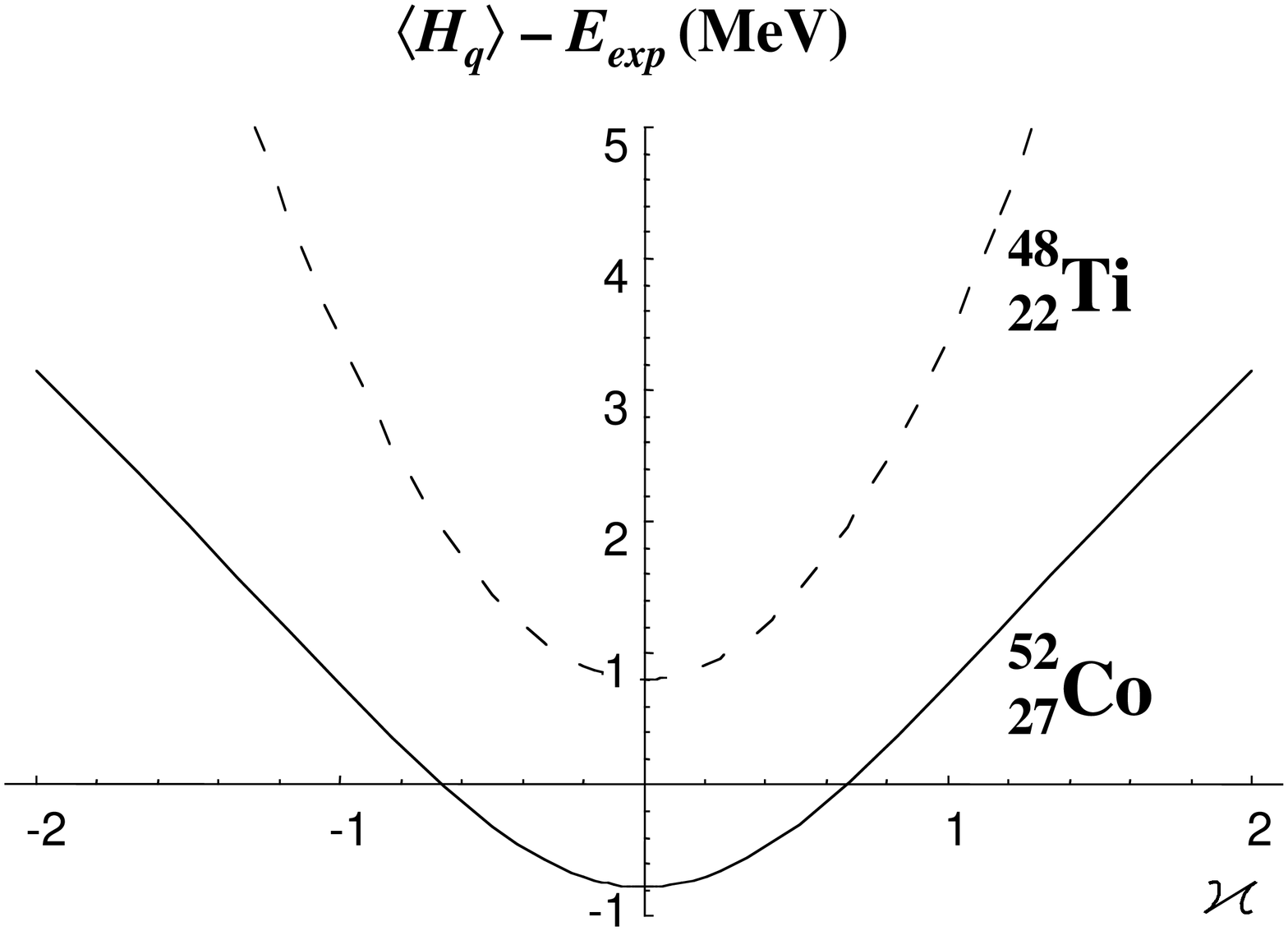,width=6cm,height=3.5cm}}}
\caption{The difference between the theoretical and the corresponding
experimental energies as a function of the deformation
parameter $\varkappa$ for a typical near-closed shell nucleus
(solid line) and for a mid-shell nucleus (dashed line). }
\label{fig:Hvskappa}
\end{figure}

\noindent\textit{3.~Novel properties of $q$-deformation}
--- To explore the physics of $q$-deformation, we fit the eigenvalues of the
deformed Hamiltonian to the relevant experimental $0^{+}$
state energies 
for groups of 36 and 100 nuclei.
This was carried out in two steps. First we determined a set of $\gamma$
parameters for the non-deformed limit 
($q=1$) 
of the theory that yielded a best
overall fit to the data.
Then another global fit was made with the $\gamma _q$ and
$q$ parameters allowed to vary. The $\gamma _q$ set that was found 
differed very
little from that of the non-deformed case. In short, varying the deformation
parameter affected the pairing strengths very little.  This observation
underscores the fact that the deformation represents something fundamentally
different, a feature that cannot be  ``mocked up'' by allowing the strengths of
the non-deformed interaction to absorb its effect.
\newline 
Once the general $\gamma $ parameters were determined, the deviation
of the predicted $0^+$ state energy from the corresponding 
experimental number,
$\left| \left\langle H_q \right\rangle -E_{\exp }\right| ^{2}$, was minimized
with respect to $q$ for each nucleus.  This procedure yielded either
two symmetric solutions for $\varkappa$ ($\left\langle H_q
\right\rangle =E_{\exp }$), that is one physical solution $|\varkappa|$,
or one value of $q$ at the minimum of the
$q$-deformed energy $\langle H_q
\rangle $ (see Fig.~\ref{fig:Hvskappa}). In the second case the
minimum occurs at the ``classical" energy ($\varkappa =0$) and its
difference 
from the experimental value can be
attributed to  the presence of other types of interactions that are not in the
model.
\newline 
The higher-order terms, which correspond to many-body interactions,
can be recognized through the expansion of the eigenvalues of the
$q$-deformed Hamiltonian (\ref{qH}).
As an example, in the
limit of identical pairing interactions, \i.e. $\suq{\pm}{2}$ subalgebra
(table~\ref{tab:su2}), and in the
$pn$ pairing limit, \i.e. $\suq{0}{2}$ limit, the ratio between the
$q$-deformed $E_q^{\pm }$ ($E_q^{pn}$) and ``classical" $E^{\pm
}$ ($E^{pn}$) energies is
\begin{align}
    R^{pp(nn)}\textstyle{ = \frac{E_q^{\pm }}{E^{\pm }} }
    =& \ 1 +
\textstyle{\frac{\varkappa^2}{6} \left( \frac{8n_{\pm
1}^2-5+4\Omega^2}{8\Omega^2}+\left( \frac{E^{\pm }}{F n_{\pm 1}}
      \right)^2 \right) + ...}, \label{eq:su2pm} \\
    R^{pn}\textstyle{ = \frac{E_q^{pn}}{E^{pn}} }
    =& \ 1 +
\textstyle{\frac{\varkappa ^2}{6}\left(\frac{n_0^2-1-4\Omega^2}{4\Omega^2}
    +\left(\frac{E^{pn}}{G n_0}\right)^2 \right) +...}, \label{eq:su2zero}
\end{align}
where the expansions include higher-order terms that may not be
negligible and the like-particles limit (\ref{eq:su2pm}) can be compared
with the earlier studies \cite{Sharma,SharmaSh}. While the quadratic coefficient
in
$R^{pp(nn)}$ is positive, the one in $R^{pn}$ is negative. This leads to a
decrease of the binding energy of the $pn$ pairs as $|\varkappa |$ increases
from zero. As the deformation parameter increases from the ``classical'' limit,
the like-particle pairing is strengthen, yielding a larger pairing
gap.

\begin{figure}[th]
\centerline{\hbox{\epsfig{figure=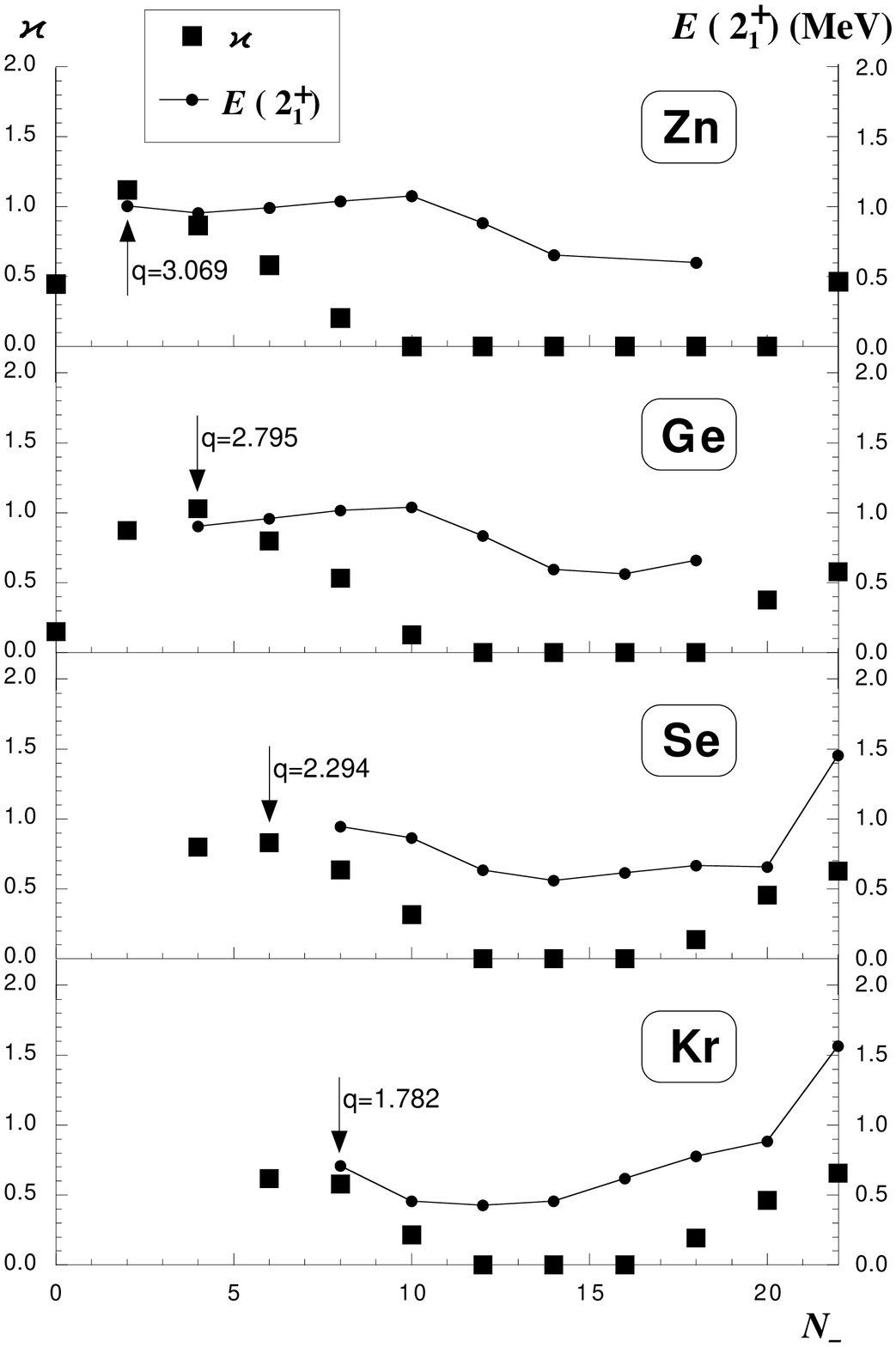,width=7cm,height=9cm}}}
\caption{Deformation parameter $\varkappa$ (symbol {\tiny $\blacksquare $})
as a function of neutron numbers ($N_-$) for various isotopes with $^{56}$Ni as
a core. The solid line is the excitation energies of the $2^+_1$ level measured
in MeV. The arrows indicate $N_+=N_-$ with the value of $q=e^{\varkappa }$. The
global  parameters are $\varepsilon=13.851$ MeV, $F/\Omega =0.296$ MeV,
$G/\Omega=0.352$ MeV, $C=0.190$ MeV, $D=-0.796$ MeV, $E/(2\Omega)=-0.489$
MeV.}
\label{fig:Ni}
\end{figure}

\noindent\textit{4.~Analysis of the role of the $q$-deformation} ---
The analysis yields solutions for
the deformation parameter $|\varkappa |$ that fall on a smooth curve that tracks
with the energy of the lowest $2^+$ states (see Fig.~\ref{fig:Ni}). 
These energies are largest near closed shells where the pairing effect
is essential for determining
the low-lying spectrum and decrease with increasing collectivity and shape
deformation. 
Similar properties are suggested for the $q$-deformation and this
result, even though  qualitative, gives some insight into the understanding of
the nature of the $q$-deformation. The observed smooth behavior of the
deformation parameter reveals its functional dependence on the model quantum
numbers.
\newline 
The many-body nature of the interaction is most important around
closed shells and the regions with $N_+ \approx N_-$.  For these nuclei the
$q$-parameter has significant values and the experimental energies can be
reproduced exactly. An interesting point is that $q$ tends to
peak for even-even nuclei when $N_+=N_-$ where strong pairing correlations are
expected (see Fig.~\ref{fig:Ni}).
\newline 
``Classical'' values of the $q$-deformation parameter ($q\approx 1$) are found
in nuclei with only one or two particle/hole pairs from a closed shell. This
is
an expected result since the number of particles is 
insufficient to sample
the effect of higher-order terms in a deformed interaction. For these 
nuclei the
non-deformed limit gives a good description.
Around
mid-shell ($N\approx 2\Omega $) the deformation adds little improvement to
the theory with the experimental values remaining close to the ``classical''
limit. This suggests that for these nuclei the pairing interaction is not
sufficient 
for their description. The results imply that even though the
$q$-parameter gives additional freedom for all the nuclei, it only improves
the model around regions of dominant pairing correlations.

A $q$-deformed extension of the $Sp(4)$ model, which is the underlying symmetry
for describing isovector ($T=1$) pairing correlations in atomic nuclei, has
been investigated.  
When compared to experimental data
the theory 
shows a smooth functional dependence of the
deformation parameter $q$ on the proton and neutron numbers, which
resembles the behavior of the lowest $2^+$ state energies. Since a
$q$-deformation of
$Sp_{q}(4)$ introduces higher-order, non-linear terms in the $np,$ $pp$ and $nn$
pairing interactions into the nuclear
Hamiltonian, the outcome suggests the presence and importance of higher-order
pairing correlations in nuclei, especially for nuclei just beyond closed shells
and with $N_+ \approx N_-$.  The results also show that the
deformation is decoupled from parameters that are used to characterize the
two-body interaction itself, which means the latter can be assigned
best-fit global values for the model space under consideration without
compromising overall quality of the theory. Moreover, the specific features of
the nuclear structure can be investigated through the use of a local $q$ value
that varies smoothly with nuclear mass number. In summary, the concept of
quantum  deformation
has been linked to the smooth behavior of physical phenomena in atomic nuclei.

This work was supported by the US National Science Foundation, Grant
Numbers 9970769 and 0140300.


\begin{thebibliography}{9}
\bibitem{Feng01}  Lianrong Dai, Feng Pan, J.P.~Draayer, J. Phys. A
\textbf{34}, 6585 (2001); 6595 (2001).

\bibitem{Jimbo}  M.~Jimbo, Lett. Math. Phys. \textbf{10}, 63 (1985);
    \textbf{11}, 247 (1986).

\bibitem{Drinfeld}  V.~Drinfeld, Proc. Int. Congress on Math, Vol. 1
    (UCPress, Berkeley: 1986).

\bibitem{Fad88}  L.D.~Faddeev, N.Y.~Reshetikhin, L.A.~Takhtajan, Algebr.
    Anal. \textbf{1}, 129 (1988).

\bibitem{Paw01}  J.~Pawelczyk and H.~Steinacker, J. High Energy Phys.
    \textbf{12}, 108 (2001).

\bibitem{Sli01}  J.K.~Slingerland and F.A.~Bais, Nucl. Phys. B \textbf{612},
    229 (2001).

\bibitem{Alg02}  A.~Algin, M.~Arik, and A.S.~Arikan, Phys. Rev. E \textbf{65},
    026140 (2002).

\bibitem{Bon00}  D.~Bonatsos et al., Phys. Rev. A \textbf{62}, 013203 (2000).

\bibitem{RRS90}  P.P.~Raychev, R.P.~Roussev, Yu.F.~Smirnov, J. Phys. G
\textbf{16}, L137 (1990).

\bibitem{BNW93}  C.~Baktash, W.~Nazarewicz, and R.~Wyss, Nucl. Phys. A
\textbf{555},
    375 (1993).

\bibitem{Sharma}  S.~Shelly~Sharma, Phys. Rev. C \textbf{46}, 904 (1992).
\bibitem{SharmaSh}  S.~Shelly~Sharma and N.K.~Sharma, Phys. Rev. C
    \textbf{62}, 034314 (2000)

\bibitem{Ker}  A.K.~Kerman, Ann. Phys. (NY) \textbf{12}, 300 (1961).

\bibitem{Rac}  G.~Racah, Phys. Rev \textbf{63}, 367 (1943).

\bibitem{Flowers}  B.H.~Flowers, Proc. Roy. Soc. (London) \textbf{A212},
    248 (1952).

\bibitem{Ros}  G.~Rosensteel and D.J.~Rowe, Phys. Rev. Lett. \textbf{38},
10 (1977); Ann. Phys. (NY) \textbf{126}, 343 (1980).

\bibitem{Helmers}  K.~Helmers, Nucl. Phys. \textbf{23}, 594 (1961).

\bibitem{Szpikowski}  B.H.~Flowers and S.~Szpikowski, Proc. Phys. Soc.
     \textbf{84}, 193 (1964).

\bibitem{Hecht}  K.T.~Hecht, Nucl. Phys. \textbf{63}, 177 (1965);
    Phys. Rev. \textbf{139}, B794 (1965); Nucl. Phys. A \textbf{102}, 11 (1967).

\bibitem{EngelLV96}  J. Engel, K. Langanke, and P. Vogel, 
Phys. Lett. {\bf B389}, 211 (1996)

\bibitem{Pang69}  S.C.~Pang, Nucl. Phys. A \textbf{128}, 497 (1969).

\bibitem{EngelPSVD97} J.~Engel et al., 
Phys. Rev. C \textbf{55}, 1781 (1997).

\bibitem{PalDJ01} Yu.V.~Palchikov, J.~Dobes, and R.V.~Jolos, Phys. Rev. C
\textbf{63}, 034320 (2001)

\bibitem{Hayashi90} T.~Hayashi, Commun. Math. Phys. \textbf{127}, 129 (1990).

\bibitem{Sel95} B.~Abdesselam, D.~Arnaudon, and A.~Chakrabarti, J. Phys. A
\textbf{28}, 3701 (1995).

\bibitem{fermRealSp4}  K.D.~Sviratcheva, A.I.~Georgieva, V.G.~Gueorguiev,
J.P.~Draayer, M.I.~Ivanov, J. Phys. A \textbf{34}, 8365 (2001).

\bibitem{KleinMarshalek} A.~Klein and E.~Marshalek, Rev. Mod. Phys.
\textbf{63}, 375 (1991).

\bibitem{Svi02}  K.D.~Sviratcheva, A.I.~Georgieva, and J.P.~Draayer,
    accepted in J.~Phys. G.

\end{thebibliography}
\end{document}